\documentclass[11pt, oneside]{article}   	
\usepackage{geometry}                		
\usepackage{graphicx}				
\usepackage{amssymb}
\usepackage{amsmath}
\usepackage{color}
\usepackage{animate}
\usepackage{soul}
\usepackage{subcaption}
\usepackage{authblk}
\usepackage{empheq}

\usepackage{graphicx}
\usepackage{cancel}
\usepackage{multicol}

\title{Modeling and Stabilizing Financial Systemic Risk Using Optimal Control Theory}
\author{Jiacheng Wu}
\affil{University of California Berkeley}
\date{}							

\begin{document}

\maketitle

\section*{Abstract}

A theoretical model of systemic-risk propagation of financial market is analyzed for stability. The state equation is an unsteady diffusion equation with a nonlinear logistic growth term, where the diffusion process captures the spread of default stress between interconnected financial entities and the reaction term captures the local procyclicality of financial stress. The stabilizing controller synthesis includes three steps: First, the algebraic Riccati equation is derived for the linearized system equation, the solution of which provides an exponentially stabilizing controller. Second, the nonlinear system is treated as a linear system with the nonlinear term as its forcing term. Based on estimation of the solutions for linearized equations and the contraction mapping theorem, unique existence of the solution for the nonlinear system equation is proved. Third, local asymptotic stability of the nonlinear system is obtained by considering the corresponding Hamilton-Jacobi equation. In both the linearized and nonlinear systems, the resulting controllers ensure that the $H^{\infty}$ norms of the mappings from disturbance to the output are less than a predefined constant. Stabilizing conditions provide a new framework of achieving system-level financial risk managing goals via the synergy of decentralized components, which offers policy-relevant insights for governments, regulators and central banks to mitigate financial crises.

\section{Introduction}

Financial systems are complex networks of financial institutions and market participants where shocks of distress can propagate through dense webs of interconnections, amplifying localized disturbances into global crises \cite{acemoglu2015systemic}. The 2008 global financial crisis and the subsequent contagion reaction underscores how seemingly uncorrelated financial entities can become deeply coupled via inter-bank exposure, linked balance-sheet and correlated asset holdings. Therefore, understanding the mechanism of risk propagation dynamics in terms of how distress spreads, amplifies, and potentially stabilizes \cite{haldane2011systemic, battiston2012debtrank}, is of great importance for policy-level risk mitigation and society-level financial security.

Traditional network-contagion model captures some features of financial systems by a network with financial institutions as nodes and exposure as weighed edge \cite{gai2010contagion,allen2000financial}. However, such models often lack an explicit treatment of continuous time and spatial propagation mechanisms, which potentially poses difficulties for application of advanced mathematical tools to model the unsteady dynamics. Borrowing the ideas from information theory and epidemiology, diffusion is a natural mechanism to model the spread of financial stress across sectors, geographies, and asset classes through liquidity flows, information channels, and overlapping portfolios. On the other hand, local dynamics such as leverage feedback, liquidity spirals and credit contagion can further amplify the distress, which is also a key component to model financial networks \cite{adrian2010liquidity,brunnermeier2009market,diamond1983bank}.

To capture these dual features of diffusion and local dynamics, the following diffusion-reaction partial differential equation is used to model the time evolution of financial stress and its propagation through financial networks,
\begin{equation}
\frac{\partial}{\partial t}s(x,t) = D_s \nabla^2 s + c_2 s \left( 1 - \frac{s}{S} \right), \label{eq:state_introduction}
\end{equation}
where $s(x,t)$ is the financial stress level at location $x$ and time $t$, $D_s$ is the diffusion coefficient represents how tightly different financial institution is connected, and $c_2$ is the growth coefficient. Note that $x$ here represents spatial dimensions beyond geographic locations, and it can also represent generic distance between financial entities in the network. The first term on the right-hand side of equation (\ref{eq:state_introduction}) describes distress propagation through financial interconnections. The second term corresponds to local amplification of distress even without contagion. $S$ is the saturated value for the distress $s(x,t)$. The state equation (\ref{eq:state_introduction}) has two fixed points $s(x,t)=0$ and $s(x,t)=S$. The equilibrium point $s(x,t)=0$ corresponds to zero distress, which is desired but unstable, while $s(x,t)=S$ corresponds to complete liquidation corresponding to financial crisis and this equilibrium is stable without intervention. Therefore, the goal is to stabilize the system equation (\ref{eq:state_introduction}) about its zero state to attenuate systemic financial distress.

Stability of linear partial differential equations is well understood using semigroup theory \cite{hille1996functional,pazy2012semigroups}. The key idea is to consider the partial differential equation as an ordinary differential equation in a functional space and prove that the corresponding spatial operator generates an exponentially stable semigroup. In the case of nonlinear systems, stability properties usually need to leverage Lyapunov indirect method \cite{lyapunov1992general,justus2008ecological,wu2016stability}. However, the theory is less well developed there, in particular for the case of partial differential state equations. Here, a novel method is proposed to handle the stabilizing controller synthesis problem for the specific nonlinear growth term in equation (\ref{eq:state_introduction}) by making use of the relationship between the nonlinear state equation and its linearized version. Optimal control theory is used to give the mathematical form of the stabilizing controller.

After formulating the control problem in Section \ref{sec:formulation}, the optimal control problem for the linearized system is considered in Section \ref{sec:linear}. Pontryagin's minimum principle \cite{pontryagin1962} is used to derive the algebraic Riccati equation, the solution of which gives the exact form of the stabilizing controller. Based on the resulting controller, it is proved that the $H^{\infty}$ norm of the mapping from the disturbance to the output is less than a preset constant. In section \ref{sec:nonlinear}, the linearized system with additional forcing terms is considered. Second, the norm estimation for the solution is obtained with respect to the norm of the forcing terms. Third, by setting the nonlinear term in the state equation (\ref{eq:state_introduction}) to be the forcing term, the existence and uniqueness of the solution for the nonlinear state equation is proved based on the contraction mapping theorem. Finally, Pontryagin's minimum problem is used again for the nonlinear optimal control problem to obtain the Hamilton-Jacobi equation, which is a nonlinear version of the Riccati equation. The feedback controller is synthesized based on the solution of the Hamilton-Jacobi equation, and local asymptotic stability is proved.

\section{Formulation of the control problem}\label{sec:formulation}

Consider the controlled state equation with control input $u(x,t)$ and disturbance $w(x,t)$,
\begin{equation}
\frac{\partial}{\partial t}s(x,t)=D_s \nabla^2 s+c_2 s \left( 1-\frac{s}{S} \right) + B_1 w(x,t) + B_2 u(x,t), \label{eq:state}
\end{equation}
where $s(x)\in H_0^1(\Omega)$, and $B_1$ and $B_2$ are input operators of the disturbance and control input, respectively. Note that $\Omega$ is the control domain, which in this case is the landscape of distributed financial entities.  The Neumann boundary conditions are
\begin{equation}
\nabla s \left.\frac{x}{|x|}(x,t) ~\right|_{x\in \Gamma}=0,
\end{equation}
where $\Gamma$ is the boundary of the domain $\Omega$. The homogeneous Neumann boundary condition enforces that there is no distress flux across the boundary. The initial condition is
\begin{equation}
s(x,0)=s_0(x),
\end{equation}
where $s_0$ represents the initial level of financial stress.

The partial differential system equation (\ref{eq:state}) may be considered to an ordinary differential equation in the Sobolev space $H_0^1(\Omega)$, with $w(t)$ as the disturbance and $u(t)$ as the control input according to
\begin{equation}
\frac{d}{dt}s(t)+D_s As(t)+F(s(t))=B_1 w(t)+B_2 u(t), \label{eq:nonlinear_state}
\end{equation}
where
\begin{equation}
As:=-\nabla^2 s,
\end{equation}
\begin{equation}
F(s(t))=F_0 s(t)+F_N(s(t))=-c_2 s+\frac{c_2}{S}s^2. \label{eq:nonlinear_terms}
\end{equation}
$F_0 s(t)$ is the linear part of the function $F(s(t))$, and $F_N(s(t))$ is the nonlinear part. Therefore, the linearized state equation about the state $s(t)=0$ is
\begin{equation}
\frac{d}{dt}s(t)+D_s As(t)+F_0 s(t)=B_1 w(t)+B_2 u(t).
\end{equation}
Note that $B_1$ and $B_2$ are assumed to be bounded linear operators, and the operator $A$ with corresponding homogeneous Neumann boundary condition is a self-adjoint operator.

\section{Optimal control for the linearized system}\label{sec:linear}

Before considering the full nonlinear state equation, a stability analysis of the corresponding linearized equation is carried out, the results of which will give one of the conditions of the controller synthesis for the nonlinear system. 

\subsection{Algebraic Riccatti equation}

The linearized system is given by
\begin{equation}
\frac{d}{dt}s(t)+D_s As(t)+F_0 s(t)=B_1 w(t)+B_2 u(t), \label{eq:linear_state}
\end{equation}
\begin{equation}
y(t)=Cs(t)+Du(t) \label{eq:linear_output}.
\end{equation}
Here, $y(t)$ denotes the output of the system, and $C$ and $D$ are bounded linear operators.
We construct the LQ optimal control problem for the linear system as follows
\begin{equation}
\sup\limits_{w}\inf\limits_{u} \frac{1}{2}\int\limits_{0}^{+\infty}(|y(t)|^2-\gamma^2|w(t)|^2)dt~~~
\mbox{subject to~equations (\ref{eq:linear_state}) and (\ref{eq:linear_output})}, \label{cost_functional_linear}
\end{equation}
where $\gamma$ is a weight coefficient.  Roughly speaking, the cost functional minimizes the influence from the disturbance on the output.  Without loss of generality, we can set $|y(t)|^2=|Cs(t)|^2+|u(t)|^2$.

Now construct Pontryagin's Hamiltonian for the system (\ref{eq:linear_state}) and (\ref{eq:linear_output}) as
\begin{equation}
H(s,p,u,w)=\langle B_1w+B_2 u-D_s As-F_0 s, p\rangle-\frac{1}{2}(|Cs|^2+|u|^2-\gamma^2|w|^2),
\label{eq:pont_ham}
\end{equation}
where $\left< \cdot, \cdot \right>$ represents the inner product, and $p(t)$ is the Lagrange multiplier used to adjoin the state equation (\ref{eq:linear_state}) to the cost functional (\ref{cost_functional_linear}).  
Based on Pontryagin's minimum principle, an extremum is reached when
\begin{eqnarray}
 \frac{\partial H}{\partial u} = 0   & \displaystyle \frac{\partial H}{\partial w} = 0,
\end{eqnarray}
which yields
\begin{eqnarray}
 u=B_2^*p,  & \displaystyle w=-\frac{1}{\gamma^2}B_1^*p, \label{eq:linear_u_w}
\end{eqnarray}
respectively, where asterisks denote adjoint operators.  The state equation and adjoint equation can be obtained from
\begin{equation}
\frac{\partial H}{\partial p}= \dot{s},~~~~\frac{\partial H}{\partial s}= -\dot{p},
\end{equation}
respectively.  Therefore, the Hamiltonian system is formulated as
\begin{eqnarray}
&&\dot{s}+D_s As+F_0 s =B_1w+B_2 u, \nonumber \\
&&\dot{p}-D_s Ap-F_0^*p = C^*Cs, \nonumber \\
&& s(0)=s_0,~~p(+\infty)=0. \label{eq:linear_system}
\end{eqnarray}
Because the system is linear, assume the invariant manifolds of the Hamiltonian system are such that
\begin{equation}
p(t)=-Ps(t), \label{eq:linear_manifold}
\end{equation}
where $P>0$ and is a self-adjoint bounded operator. 

Because the system is autonomous, and the terminal time is not specified, $H(s,p)=0$ for $t>0$.  Substituting equations (\ref{eq:linear_u_w}) and (\ref{eq:linear_manifold}) into $H(s,p)=0$ from equation (\ref{eq:pont_ham}) yields
\begin{equation}
\left\langle P(D_sA+F_0)s-\frac{1}{2}C^*Cs+\frac{1}{2}PB_2B_2^*Ps-\frac{1}{2\gamma^2}PB_1B_1^*Ps,s\right\rangle=0,
\end{equation}
which gives the algebraic Riccati equation
\begin{equation}
P(D_sA+F_0)-\frac{1}{2}C^*C+\frac{1}{2}PB_2B_2^*P-\frac{1}{2\gamma^2}PB_1B_1^*P=0. \label{eq:Riccati}
\end{equation}
The solution for $P > 0$ defines the mathematical form of the stabilizing feedback controller $u = -B_2^* P s$.

\subsection{Stability of the linearized system with no disturbances}

In order to prove stability of the linearized system with the no disturbances, that is, $w=0$, first construct the Lyapunov function from the solution to the algebraic Riccati equation as
\begin{equation}
V(s(t))=\frac{1}{2} \left< s, Ps \right> \geq 0.
\end{equation}
Taking the time derivative of $V(s(t))$ along the trajectory of the linearized system gives
\begin{eqnarray}
\frac{dV(s(t))}{dt}&=&\left\langle\dot{s}, \frac{dV(s)}{ds}\right\rangle \nonumber \\
&=&\langle B_2 u-(D_sA+F_0)s,Ps\rangle \nonumber \\
&=&\langle PB_2u-P(D_sA+F_0)s,s\rangle.
\end{eqnarray}
Substituting $u=B_2^*p=-B_2^*Ps$ and the algebraic Riccati equation (\ref{eq:Riccati}) into this equation requires
\begin{eqnarray}
\frac{dV(s(t))}{dt}&=&\left\langle -\frac{1}{2}PB_2B_2^*Ps-\frac{1}{2\gamma^2}PB_1B_1^*Ps-\frac{1}{2}C^*Cs, s   \right\rangle \nonumber \\
&=&-\frac{1}{2}|B_2^*Ps|^2-\frac{1}{2\gamma^2}|B_1^*Ps|^2-\frac{1}{2}|Cs|^2 \nonumber \\
&\leq&-m|s|^2,
\end{eqnarray}
where $m>0$. This result demonstrates the global exponential stability of the linearized system with $w=0$
via the state feedback controller $u=-B_2^*Ps$, and that the operator $-(D_sA+F_0+B_2 B_2^*P)$ generates an exponentially stable semigroup.  This result ensures the unique existence of the solution for the linearized system and will be used later to prove asymptotic stability of the nonlinear system.

\subsection{Boundedness of the $H^{\infty}$ norm of the mapping from $w$ to $y$}

Once stability is assured for the linearized system with no disturbances, it is sought to demonstrate that the influence from disturbances on the output is smaller that $\gamma$. Consider the linearized system with zero initial condition
\begin{equation}
\frac{d}{dt}s(t)+D_s As(t)+F_0 s(t)=B_1 w(t)+B_2 u(t),~~s(0)=0. \label{eq:linear_w}
\end{equation}
Taking the time derivative of the Lyapunov function $V(s(t))=\frac{1}{2}\langle s,Ps\rangle$ along the trajectory of equation (\ref{eq:linear_w}) gives
\begin{eqnarray}
\frac{dV(s(t))}{dt}&=&\left\langle\dot{s}, Ps\right\rangle \nonumber \\
&=&\langle B_1 w+B_2 u-(D_sA+F_0)s,Ps\rangle \nonumber \\
&=&\langle PB_1 w +PB_2 u-P(D_sA+F_0)s,s\rangle.
\end{eqnarray}
Substituting $u=B_2^*p=-B_2^*Ps$ and the algebraic Riccati equation (\ref{eq:Riccati}) into this equation lead to
(note that we do not set $w=-\frac{1}{\gamma^2}B_1^*p$)
\begin{eqnarray}
\frac{dV(s(t))}{dt}&=&\left\langle PB_1 w-\frac{1}{2}PB_2B_2^*Ps-\frac{1}{2\gamma^2}PB_1B_1^*Ps-\frac{1}{2}C^*Cs, s   \right\rangle \nonumber \\
&=&\langle PB_1 w,s\rangle-\frac{1}{2}|B_2^*Ps|^2-\frac{1}{2\gamma^2}|B_1^*Ps|^2-\frac{1}{2}|Cs|^2 \nonumber \\
&=&-\frac{1}{2}\gamma^2|w|^2+\langle w,B_1^*Ps \rangle-\frac{1}{2\gamma^2}|B_1^*Ps|^2+\frac{1}{2}\gamma^2|w|^2-\frac{1}{2}|Cs|^2-\frac{1}{2}|u|^2 \nonumber \\
&=&-\frac{1}{2}\gamma^2\left|w-\frac{1}{\gamma^2}B_1^*Ps  \right|^2+\frac{1}{2}\gamma^2|w|^2-\frac{1}{2}|Cs|^2-\frac{1}{2}|u|^2 \nonumber \\
&\leq& \frac{1}{2}\gamma^2|w|^2-\frac{1}{2}|Cs|^2-\frac{1}{2}|u|^2.
\end{eqnarray}
Integrate the equation above from $t=0$ to $t=+\infty$,
\begin{equation}
V(s(+\infty))-V(s_0)<\int\limits_0^{+\infty}\left(\frac{1}{2}\gamma^2|w|^2-\frac{1}{2}|Cs|^2-\frac{1}{2}|u|^2\right)dt.
\end{equation}
For the initial condition $s_0=0$, $V(s_0)=0$. Also, because $V(s(+\infty))\geq 0$, it can be concluded that
\begin{equation}
\int\limits_0^{+\infty}(|Cs|^2+|u|^2)dt<\gamma^2\int\limits_0^{+\infty}|w|^2dt.
\end{equation}
This demonstrates that the $H^{\infty}$ norm of the mapping from the disturbance $w$ to the output $y$ is less than $\gamma$, which is the weight coefficient in the cost functional (\ref{cost_functional_linear}).

\section{Optimal control for the nonlinear system} \label{sec:nonlinear}

In the linear case, the fact that the operator $-(D_sA+F_0+B_2 B_2^*P)$ generates an exponentially stable semigroup ensures the unique existence of the solution for the linearized state equation. However, this is generally not the case for the nonlinear situation. Therefore, the contraction mapping theorem will be used to prove existence and uniqueness of the solution for the nonlinear state equation. 

\subsection{Construct a linearized system with a forcing term and no disturbances}

The idea here is to treat the nonlinear term as an additional forcing term in the linearized system. Adding a forcing term $f(t)$
to the linearized state equation considered in the previous section gives
\begin{equation}
\frac{d}{dt}s(t)+D_s As(t)+F_0 s(t)=B_2 u(t) + f(t). \label{eq:linear_force_state}
\end{equation}
Consider the following optimal control problem
\begin{equation}
\inf\limits_{u}\frac{1}{2}\int\limits_0^{+\infty}(|Cs|^2+|u|^2+2\langle g,s\rangle) dt ~~~
\mbox{subject to~equation}~ (\ref{eq:linear_force_state}) \label{optimal_control_linear_force},
\end{equation}
where $g(t)$ will be a forcing term in the differential adjoint equation.

Constructing the Pontryagin Hamiltonian yields
\begin{equation}
H(s,p,u)=\langle B_2 u-D_s As-F_0 s+f,p\rangle-\frac{1}{2}(|Cs|^2+|u|^2+2\langle g,s\rangle).
\end{equation}
Note that the function $g(t)$ will show up as a forcing term in the adjoint equation.

From Pontryagin's minimum principle, an extremum is reached when
\begin{equation}
 \frac{\partial H}{\partial u} = 0;
\end{equation}
thus
\begin{equation}
 u=B_2^*p,
\end{equation}
and the Hamiltonian system corresponding to the optimal control problem (\ref{optimal_control_linear_force}) is
\begin{eqnarray}
&&\dot{s}+D_s As+F_0 s =B_2 u+f, \nonumber \\
&&\dot{p}-D_s Ap-F_0^*p = C^*Cs +g ,  \nonumber \\
&&s(0)=s_0,~~p(+\infty)=0.   \label{eq:linear_force_system}
\end{eqnarray}
The Hamiltonian system is written in the following form
\begin{eqnarray}
&&\dot{s}+(D_s A+F_0 +B_2B_2^*P)s=f, \nonumber \\
&&\dot{p}-(D_s A+F_0+B_2B_2^*P)^*p = C^*Cs-PB_2B_2^*p +g.   \label{eq:linear_force_system2}
\end{eqnarray}
Because it has been proved that the operator $-(D_s A+F_0 +B_2B_2^*P)$ generates an exponentially stable semigroup, then $\exists~ a>0$ such that
\begin{equation}
|s(t)|\leq |s_0|e^{-at}+\int\limits_0^{t}e^{-a(t-r)}|f(r)|dr, \label{ineq:estimate_s}
\end{equation}
and
\begin{equation}
|p(t)|\leq C\int\limits_t^{+\infty}e^{-a(r-t)}(|Cs(r)|+|B_2^*p(r)|+|g(r)|)dr. \label{ineq:estimate_p}
\end{equation}
Let us first calculate the estimate for the norm of $s(t)$ in the space $L^2(0, \infty; L^2(\Omega))$, denoted by $\|\cdot\|$,
by integrating the square  of equation (\ref{ineq:estimate_s}) from $0$ to $+\infty$
\begin{eqnarray}
\|s(t)\|^2&=&\int\limits_0^{+\infty}|s(t)|^2dt \nonumber \\
&\leq& |s_0|^2\int\limits_0^{+\infty}e^{-2at}dt +\left\| e^{-at}*|f(t)| \right\|^2 \nonumber \\
&\leq& |s_0|^2\int\limits_0^{+\infty}e^{-2at}dt +\left(\int\limits_0^{+\infty}\left| e^{-at}\right|dt\right)^2    \left\| f(t)\right\|^2 \nonumber \\
&\leq& C\left( |s_0|^2+\|f(t)\|^2 \right). \label{ineq:estimate_timenorm_s}
\end{eqnarray}
The derivation of this inequality has made use of Young's inequality for convolution.  Similarly, from equation (\ref{ineq:estimate_p}),
the estimate for the norm of $s(t)$ in the space $L^2(0, \infty; L^2(\Omega))$ is
\begin{eqnarray}
\|p(t)\|^2&\leq&C\left\| |Cs|+ |B_2^*p|+|g|  \right\|^2 \nonumber \\
&\leq& C\left(  \int\limits_0^{+\infty}\left(|Cs|^2+|B_2^*p|^2\right)dt +\|g\|^2   \right). \label{ineq:estimate_timenorm_p}
\end{eqnarray}
From equation (\ref{eq:linear_force_system2}),
\begin{eqnarray}
|Cs|^2+|B_2^*p|^2&=&\langle C^*Cs,s\rangle+\langle B_2 B_2^*p,p\rangle \nonumber \\
&=&\left\langle\dot{p}-(D_sA+F_0)^*p-g,s\right\rangle+\left\langle B_2 B_2^*p,p\right\rangle \nonumber \\
&=&\left\langle \dot{p}, s \right\rangle-\langle g,s\rangle+\left\langle p, -(D_sA+F_0)s+B_2B_2^*p\right\rangle \nonumber \\
&=&\left\langle \dot{p}, s \right\rangle-\langle g,s\rangle+\left\langle p, \dot{s}-f \right\rangle \nonumber \\
&=&\frac{d}{dt}\langle p,s\rangle-\langle g,s\rangle-\langle p,f\rangle.
\end{eqnarray}
Integrating this from $0$ to $+\infty$, it is found that
\begin{eqnarray}
\int\limits_0^{+\infty}\left( |Cs|^2+|B_2^*p|^2 \right)dt&=&-\langle p(0),s(0)\rangle-\int\limits_0^{\infty}\langle g,s\rangle dt-\int\limits_0^{\infty}\langle p,f\rangle dt \nonumber\\
&\leq& \langle |p(0)|,|s_0|\rangle+\int\limits_0^{+\infty}|\langle g,s\rangle|dt+\int\limits_0^{+\infty}|\langle p,f\rangle|dt.
\end{eqnarray}
Substituting $t=0$ into equation (\ref{ineq:estimate_p}) gives the following estimate for $|p(0)|$:
\begin{equation}
|p(0)|\leq C\int\limits_0^{+\infty}e^{-ar}(|Cs(r)|+|B_2^*p(r)|+|g(r)|)dr.
\end{equation}
Therefore,
\begin{eqnarray}
\int\limits_0^{+\infty}\left( |Cs|^2+|B_2^*p|^2 \right)dt&\leq& C\left\langle\int\limits_0^{+\infty}e^{-ar}(|Cs(r)|+|B_2^*p(r)|+|g(r)|)dr ,|s_0|\right\rangle \nonumber \\
&&+\int\limits_0^{+\infty}|\langle g,s\rangle|dt+\int\limits_0^{+\infty}|\langle p,f\rangle|dt \nonumber \\
&\leq& C\int\limits_0^{+\infty}\left\langle  |Cs(r)|+|B_2^*p(r)|+|g(r)| , e^{-ar}|s_0|\right\rangle dr \nonumber \\
&&+\int\limits_0^{+\infty}|\langle g,s\rangle|dt+\int\limits_0^{+\infty}|\langle p,f\rangle|dt. \label{ineq:Cs_Bp}
\end{eqnarray}
Applying Young's inequality with $\epsilon$
\begin{equation}
k_1 k_2\leq \frac{1}{2\epsilon} k_1^2+\frac{\epsilon}{2}k_2^2
\end{equation}
to the three terms on the right hand side of inequality (\ref{ineq:Cs_Bp}) yields
\begin{eqnarray}
\int\limits_0^{+\infty}\left( |Cs|^2+|B_2^*p|^2 \right)dt
&\leq& \epsilon \int\limits_0^{+\infty}\left( |Cs|+|B_2^*p|+|g| \right)^2dt+C_{\epsilon}|s_0|^2 \nonumber \\
&&+\epsilon\left( \|s\|^2+\|p\|^2 \right)+C_{\epsilon}\left( \|g\|^2+\|f\|^2 \right) \nonumber\\
&\leq & \epsilon \left(  \int\limits_0^{+\infty}\left( |Cs|^2+|B_2^*p|^2 \right)dt + \|s\|^2+\|p\|^2 \right) \nonumber \\
&&+C_{\epsilon}\left( |s_0|^2+\|g\|^2+\|f\|^2 \right).
\end{eqnarray}
This is equivalent to
\begin{equation}
\int\limits_0^{+\infty}\left( |Cs|^2+|B_2^*p|^2 \right)dt \leq
\epsilon \left(  \|s\|^2+\|p\|^2 \right)+C_{\epsilon}\left( |s_0|^2+\|g\|^2+\|f\|^2 \right).
\end{equation}
Substituting this inequality into inequality (\ref{ineq:estimate_timenorm_p}) yields
\begin{equation}
\|p(t)\|^2\leq \epsilon \left(  \|s\|^2+\|p\|^2 \right)+C_{\epsilon}\left( |s_0|^2+\|g\|^2+\|f\|^2 \right).
\end{equation}
Now combine with inequality (\ref{ineq:estimate_timenorm_s}) to obtain
\begin{equation}
\|s(t)\|+\|p(t)\|^2\leq \epsilon \left(  \|s\|^2+\|p\|^2 \right)+C_{\epsilon}\left( |s_0|^2+\|g\|^2+\|f\|^2 \right),
\end{equation}
which is equivalent to
\begin{equation}
\|s(t)\|^2+\|p(t)\|^2\leq C\left( |s_0|^2+\|g\|^2+\|f\|^2 \right).
\end{equation}
Then, with some algebraic manipulations, it can be concluded that
\begin{equation}
\|s(t)\|+\|p(t)\|\leq C\left( |s_0|+\|g\|+\|f\| \right). \label{ineq:s_p}
\end{equation}
This estimation is of great significance because it enables us to bound the norm of the solution by the norm of the initial condition and the additional forcing term. By choosing the forcing term in a special form, the nonlinear state equation can be readily connected to the linearized state equation.

\subsection{Uniqueness and existence of the solution of the nonlinear system with no disturbances}

Now consider the optimal control problem for the original nonlinear system (\ref{eq:nonlinear_state}) once again and define the cost functional in the form (\ref{cost_functional_linear}),
then the optimal control problem is formulated as
\begin{eqnarray}
\sup\limits_{w}\inf\limits_{u}&&J(s_0)=\frac{1}{2}\int\limits_0^{+\infty}\left( |Cs|^2+|u|^2-\gamma^2|w|^2 \right)dt, \nonumber \\
\mbox{subject to}&&\frac{d}{dt}s(t)+D_s As(t)+F(s(t))=B_1 w(t)+B_2 u(t),  \label{eq:optimalcontrol_nonlinear}
\end{eqnarray}
where $F(s(t))$ is given by (\ref{eq:nonlinear_terms}), which includes the nonlinear term.

Once again the Pontryagin Hamiltonian is constructed in the following way
\begin{equation}
H(s,p,u,w)=\left\langle B_1 w+B_2 u-D_sAs-F(s),p \right\rangle-\frac{1}{2}\left( |Cs|^2+|u|^2-\gamma^2|w|^2 \right).
\end{equation}
Based on Pontryagin's minimum principle, an extremum is reached, as in equation (\ref{eq:linear_u_w}), when
\begin{eqnarray}
 \frac{\partial H}{\partial u} = 0   & \displaystyle \frac{\partial H}{\partial w} = 0,
\end{eqnarray}
which yields
\begin{eqnarray}
 u=B_2^*p, & \displaystyle w=-\frac{1}{\gamma^2}B_1^*p, \label{eq:nonlinear_Hu_Hw}
\end{eqnarray}
respectively.  The corresponding Hamiltonian system is now
\begin{eqnarray}
&&\dot{s}+D_s As+F( s) =B_1w+B_2 u, \nonumber \\
&&\dot{p}-D_s Ap-(\nabla F(s))^*p = C^*Cs,  \nonumber \\
&&s(0)=s_0,~~p(+\infty)=0.   \label{eq:nonlinear_system}
\end{eqnarray}
Recall that $F(s)=F_0 s+F_N(s)$ in the nonlinear case; therefore, this becomes
\begin{eqnarray}
&&\dot{s}+D_s As+F_0 s =B_2 u-F_N(s), \nonumber \\
&&\dot{p}-D_s Ap-F_0^*p = C^*Cs +\left(\nabla F_N(s)\right)^* p,  \nonumber \\
&&s(0)=s_0,~~p(+\infty)=0.
\end{eqnarray}
This is the same as the linear system with forcing (\ref{eq:linear_force_system}), but now with
\begin{equation}
f=-F_N(s)=\frac{c_2}{S}s^2,~~g=\left(\nabla F_N(s)\right)^* p=\frac{2c_2}{S}sp.
\end{equation}
Let us define the mapping from the solution pair $(s, p)$ to the forcing terms $(f, g)$ through the nonlinear term as $\mathcal{M}_1$.  Similarly, the mapping from the forcing terms $(f, g)$ to the solution pair $(s, p)$ through the linearized system with forcing terms (\ref{eq:linear_force_system}) is $\mathcal{M}_2$.  The composite mapping is denoted by $\mathcal{M}_2\circ\mathcal{M}_1$.  If $(s,p)$ is a solution to the nonlinear system, then $(s,p) = \mathcal{M}_2\circ\mathcal{M}_1 (s,p)$.  Then if the composite mapping $\mathcal{M}_2\circ\mathcal{M}_1$ is a contraction, the solution $(s, p)$ uniquely exists.

Let $\Sigma_{\mu}$ be a subset of $H\times H$ defined as
\begin{equation}
\Sigma_{\mu}=\{(s,p)\left| \|s\|+\|p\|\leq \mu  \right.\}.
\end{equation}
Then for $(s,p)\in \Sigma_{\mu}$, we have
\begin{eqnarray}
\|f\|+\|g\|&=&\left\|\frac{c_2}{S}s^2\right\|+\left\|\frac{2c_2}{S}sp\right\| \nonumber \\
&\leq&\frac{2c_2}{S}\|s\|\left( \|s\|+\|p\| \right) \nonumber \\
&\leq&\frac{2c_2}{S}\mu^2.
\end{eqnarray}
From the inequality (\ref{ineq:s_p}), it is observed that
\begin{equation}
\|s\|+\|p\| \leq C\left( |s_0|+\|f\|+\|g\| \right)\leq   |s_0|+\frac{2c_2}{S}\mu^2.
\end{equation}
In order to make $\Sigma_{\mu}$ invariant with respect to the composite mapping $\mathcal{M}_2\circ\mathcal{M}_1$, it is necessary to set
\begin{equation}
|s_0|+\frac{2c_2}{S}\mu^2\leq \mu.
\end{equation}
Therefore, if
\begin{equation}
|s_0|<\frac{S}{8c_2C^2},
\end{equation}
then
\begin{equation}
\frac{S-\sqrt{S^2-8c_2C^2S|s_0|}}{4c_2C}\leq\mu\leq\frac{S+\sqrt{S^2-8c_2C^2S|s_0|}}{4c_2C},
\end{equation}
and $\Sigma_{\mu}$ is invariant with respect to $\mathcal{M}_2\circ\mathcal{M}_1$.

As one can see, if $(s, p)$ is a solution to the nonlinear system (\ref{eq:nonlinear_system}), then $(s, p)$ is a fixed point of the composite mapping $\mathcal{M}_2\circ\mathcal{M}_1$. Hence, in order to show that the nonlinear system (\ref{eq:nonlinear_system}) has a unique solution, it is necessary to prove that $\mathcal{M}_2\circ\mathcal{M}_1$ is a contraction mapping.

To do so, assume there are two different groups of forcing terms $(f_1, g_1)$ and $(f_2, g_2)$, and the corresponding  solutions of (\ref{eq:linear_force_system}) are $(s_1,g_1)$ and $(s_2, g_2)$, which satisfy the following systems of equations
\begin{eqnarray}
&&\dot{s_1}+D_s As_1+F_0 s_1 =B_2 B_2^*p_1+f_1, \nonumber \\
&&\dot{p_1}-D_s Ap_1-F_0^*p_1 = C^*Cs_1 +g_1,  \nonumber \\
&&s_1(0)=s_0,~~p_1(+\infty)=0,
\end{eqnarray}
and
\begin{eqnarray}
&&\dot{s_2}+D_s As_2+F_0 s_2 =B_2 B_2^*p_2+f_2, \nonumber \\
&&\dot{p_2}-D_s Ap_2-F_0^*p_2 = C^*Cs_2 +g_2, \nonumber \\
&&s_2(0)=s_0,~~p_2(+\infty)=0,
\end{eqnarray}
respectively. By linearity of the above equations, superposition leads to
\begin{eqnarray}
&&\frac{d}{dt}(s_1-s_2)+D_s A(s_1-s_2)+F_0 (s_1-s_2) =B_2 B_2^*(p_1-p_2)+(f_1-f_2), \nonumber \\
&&\frac{d}{dt}(p_1-p_2)-D_s A(p_1-p_2)-F_0^*(p_1-p_2) = C^*C(s_1-s_2) +(g_1-g_2),  \nonumber \\
&&(s_1-s_2)(0)=0,~~(p_1-p_2)(+\infty)=0.
\end{eqnarray}
Applying relation (\ref{ineq:s_p}), requires that
\begin{equation}
\|s_1-s_2\|+\|p_1-p_2\|\leq C\left( \|g_1-g_2\|+\|f_1-f_2\| \right),  \label{ineq:map_M_2}
\end{equation}
which demonstrates that the mapping $\mathcal{M}_2$ is Lipschitz continuous.

Now consider the mapping $\mathcal{M}_1$ from $(s, p)$ to $(f, g)$, for which
\begin{eqnarray}
\|f_1-f_2\|+\|g_1-g_2\|&=&\left\|\frac{c_2}{S}\left( s_1^2-s_2^2 \right)\right\|+\left\| \frac{2c_2}{S}\left( s_1 p_1-s_2 p_2 \right) \right\| \nonumber \\
&\leq& \frac{c_2}{S}\left( \|s_1\|+\|s_2\| \right) \|s_1-s_2\|  \nonumber \\
&&+ \frac{2c_2}{S}\| s_1 p_1-s_1 p_2+s_1 p_2-s_2 p_2 \| \nonumber \\
&\leq&\frac{c_2}{S}\left( \|s_1\|+\|s_2\|+2\|p_2\| \right)\|s_1-s_2\| \nonumber \\
&&+\frac{2c_2}{S}\|s_1\| \|p_1-p_2\|.
\end{eqnarray}
If the mapping is considered to be within $\Sigma_{\mu}$, that is
\begin{equation}
\|f_1-f_2\|+\|g_1-g_2\| \leq \frac{4c_2\mu}{S}\left( \|s_1-s_2\|+\|p_1-p_2\| \right),
\end{equation}
then this relation can be combined with (\ref{ineq:map_M_2}) to yield
\begin{equation}
\|s_1-s_2\|+\|p_1-p_2\|\leq \frac{4c_2 C\mu}{S}\left( \|s_1-s_2\|+\|p_1-p_2\| \right).
\end{equation}
Therefore, for the Lipschitz constant to be less than one, we set
\begin{eqnarray}
\frac{4c_2 C\mu}{S}<1,
\end{eqnarray}
which requires that
\begin{eqnarray}
\mu<\frac{S}{4c_2C}~,
\end{eqnarray}
then the composite mapping $\mathcal{M}_2\circ\mathcal{M}_1$ is a contraction mapping.
This ensures the existence and uniqueness of the solution $(s,p)$ to the nonlinear system (\ref{eq:nonlinear_system}), which is a fixed point of $\mathcal{M}_2\circ\mathcal{M}_1$.

\subsection{Conditions for the solution to the Hamilton-Jacobi equation for the nonlinear system}

Thus far, the contraction mapping theorem has been used to prove existence and uniqueness of the solution for the controlled nonlinear system equation.  Next, the mathematical form of the stabilizing controller for the nonlinear system is obtained via the Hamilton-Jacobi equation.

Recall that the Hamiltonian for the nonlinear system (\ref{eq:nonlinear_system}) is
\begin{equation}
H(s,p,u,w)=\left\langle B_1 w+B_2 u-D_sAs-F(s),p \right\rangle-\frac{1}{2}\left( |Cs|^2+|u|^2-\gamma^2|w|^2 \right).
\end{equation}
Because the system is nonlinear, assume that the mapping from $s_0$ to $p(0)$ has the nonlinear form
\begin{equation}
p(0)=-G(s_0).
\end{equation}
Because the uniqueness of the solution for the nonlinear system (\ref{eq:nonlinear_system}) has been proved,
the invariant manifold of the Hamiltonian system also takes the same nonlinear form according to
\begin{equation}
p(t)=-G(s(t)).
\end{equation}
The terminal time of the optimal control problem (\ref{eq:optimalcontrol_nonlinear}) is not specified; therefore, we have
\begin{equation}
H(s, p, u, w)=0.
\end{equation}
Substituting the invariant manifolds gives,
\begin{eqnarray}
H(s, p,u,w)&=&\left\langle \frac{1}{\gamma^2}B_1B_1^*G(s(t))-B_2B_2^*G(s(t))-D_sAs(t)-F(s(t)),-G(s(t)) \right\rangle  \nonumber \\
&&-\frac{1}{2}\left( |Cs(t)|^2+|B_2^*G(s(t))|^2-\frac{1}{\gamma^2}|B_1^*G(s(t))|^2 \right) \nonumber \\
&=&\left\langle D_sAs+F(s), G(s) \right\rangle-\frac{1}{2\gamma^2}|B_1^*G(s)|^2+\frac{1}{2}|B_2^*G(s)|^2-\frac{1}{2}|Cs|^2.
\end{eqnarray}
Therefore, the Hamilton-Jacobi equation for the nonlinear mapping $G(\cdot)$ is
\begin{equation}
\left\langle D_sAs+F(s), G(s) \right\rangle-\frac{1}{2\gamma^2}|B_1^*G(s)|^2+\frac{1}{2}|B_2^*G(s)|^2-\frac{1}{2}|Cs|^2=0 \label{eq:HJE1}.
\end{equation}
From equations (\ref{eq:nonlinear_Hu_Hw}) and (\ref{eq:nonlinear_system}), if $s=0$, then $p=0$, which yields
\begin{equation}
G(0)=0.
\end{equation}

To see the relationship between the Hamilton-Jacobi equation (\ref{eq:HJE1}) and the algebraic Riccati equation (\ref{eq:Riccati}), first consider the relationship between the cost function $J(s_0)$ and the nonlinear mapping $G$.
Because the cost functional $J(s_0)$ can be treated as a function of the initial condition $s_0$,
consider how perturbation of the initial condition will influence the solution of the nonlinear Hamiltonian system (\ref{eq:nonlinear_system}).

Assume $\delta$ is a small perturbation on the initial condition of (\ref{eq:nonlinear_system}), that is, $(s+s_{\delta})(0)=s_0+\delta$, and the solution becomes
$(s+s_{\delta}, p+p_{\delta})$. Both $s_{\delta}$ and $p_{\delta}$ are on the scale of $O(\delta)$.
\begin{eqnarray}
&&\frac{d}{dt}(s+s_{\delta})+D_s A(s+s_{\delta})+F( s+s_{\delta}) =\left( B_2 B_2^*-\frac{1}{\gamma^2}B_1B_1^*\right)(p+p_{\delta}), \nonumber \\
&&\frac{d}{dt}(p+p_{\delta})-D_s A(p+p_{\delta})-(\nabla F(s+s_{\delta}))^*(p+p_{\delta}) = C^*C(s+s_{\delta}),  \nonumber \\
&&(s+s_{\delta})(0)=s_0+\delta,~~(p+p_{\delta})(+\infty)=0.
\end{eqnarray}
Canceling the $O(1)$ terms by substituting (\ref{eq:nonlinear_system}) and neglecting $O(\delta^2)$ terms leaves the following equations for the perturbations
\begin{eqnarray}
&&\dot{s_{\delta}}+D_sAs_{\delta}+\nabla F(s) s_{\delta}=\left( B_2 B_2^*-\frac{1}{\gamma^2}B_1B_1^*\right)p_{\delta}, \nonumber \\
&&\dot{p_{\delta}}-D_sAp_{\delta}-\left( \nabla F(s) \right)^*p_{\delta}-\left( \nabla^2 F(s)s_{\delta} \right)^*p=C*Cp_{\delta}, \nonumber \\
&&s_{\delta}(0)=\delta,~~p_{\delta}(+\infty)=0.  \label{eq:delta}
\end{eqnarray}
Define the mapping from $s_0$ to $(s,p)$ as
\begin{equation}
\Phi(s_0)=(s,p).
\end{equation}
Then from the definition of $( s_{\delta}, p_{\delta})$, we have
\begin{equation}
\left\langle\nabla \Phi(s_0), \delta \right\rangle=( s_{\delta}, p_{\delta}).
\end{equation}
Because $J(s_0)=J(s(s_0),p(s_0))$,
\begin{eqnarray}
\left\langle \nabla J(s_0), \delta \right\rangle&=&\left\langle\left( \frac{\delta J}{\delta s}, \frac{\delta J}{\delta p} \right), \left( s_{\delta}, p_{\delta} \right)^T\right\rangle \nonumber \\
&=&\int\limits_0^{+\infty}\left(  \langle C^*Cs, s_{\delta}\rangle+\langle B_2B_2^*p, p_{\delta}\rangle-\left\langle \frac{1}{\gamma^2}B_1B_1^*p, p_{\delta} \right\rangle \right)dt.
\end{eqnarray}
Substituting (\ref{eq:nonlinear_system}) and (\ref{eq:delta}) yields
\begin{eqnarray}
\left\langle \nabla J(s_0), \delta \right\rangle
&=&\int\limits_0^{+\infty}\left(  \left\langle\dot{p}-D_s Ap-(\nabla F(s))^*p, s_{\delta}\right\rangle
+\left\langle B_2B_2^*p-\frac{1}{\gamma^2}B_1B_1^*p, p_{\delta} \right\rangle \right)dt \nonumber \\
&=&\int\limits_0^{+\infty}\left( \left\langle \dot{p},s_{\delta} \right\rangle -\left\langle p,D_sAs_{\delta}+(\nabla F(s))s_{\delta} \right\rangle
+\left\langle p,  B_2B_2^*p-\frac{1}{\gamma^2}B_1B_1^*p_{\delta} \right\rangle          \right)dt \nonumber \\
&=&\int\limits_0^{+\infty}\frac{d}{dt}\left\langle p, s_{\delta} \right\rangle dt \nonumber \\
&=&-\langle p(0),\delta\rangle.
\end{eqnarray}
Because $p(0)=-G(s_0)$, then we have
\begin{equation}
\nabla J(s_0)=G(s_0).
\end{equation}

Now consider the case when $s_0=0$, then the corresponding solution of the nonlinear system is $(s,p)=(0,0)$.
Substituting $(s,p)=(0,0)$ into (\ref{eq:delta}), the equations are satisfied by the perturbation about the zero solution $(s,p)=(0,0)$ as follows:
\begin{eqnarray}
&&\dot{s_{\delta}}+D_sAs_{\delta}+F_0 s_{\delta}=\left( B_2 B_2^*-\frac{1}{\gamma^2}B_1B_1^*\right)p_{\delta}, \nonumber \\
&&\dot{p_{\delta}}-D_sAp_{\delta}-F_0^*p_{\delta}=C*Cp_{\delta}, \nonumber \\
&&s_{\delta}(0)=\delta,~~p_{\delta}(+\infty)=0, \label{eq:linear_delta}
\end{eqnarray}
which have the same form as the linear Hamiltonian system (\ref{eq:linear_system}). Therefore, the invariant manifold for (\ref{eq:linear_delta})
is the same as for (\ref{eq:linear_system}), which is
\begin{equation}
p_{\delta}=-Ps_{\delta},
\end{equation}
where $P$ is the solution to the algebraic Riccati equation (\ref{eq:Riccati}). Therefore,
\begin{equation}
\left\langle\nabla \Phi(s_0), \delta \right\rangle=(s_{\delta}, p_{\delta})=(s_{\delta}, -Ps_{\delta}).
\end{equation}
On the other hand, because $\Phi(s_0)=(s,p)=(s,-G(s))$,
\begin{equation}
\left\langle\nabla \Phi(s_0), \delta \right\rangle=(s_{\delta}, -\nabla G(s) s_{\delta}).
\end{equation}
Setting $s_0=0$, the relationship between the solutions to the algebraic Riccati equation (\ref{eq:Riccati}) and Hamilton-Jacobi equation (\ref{eq:HJE1}) is given by
\begin{equation}
\nabla G(0)=P.
\end{equation}
Hence, it is concluded that the nonlinear mapping $G(\cdot)$ needs to satisfy the following four conditions:
\begin{itemize}
  \item[i)] $\left\langle D_sAs+F(s), G(s) \right\rangle-\frac{1}{2\gamma^2}|B_1^*G(s)|^2+\frac{1}{2}|B_2^*G(s)|^2-\frac{1}{2}|Cs|^2=0 \label{eq:HJE}$,
  \item[ii)] $G(0)=0$,
  \item[iii)] $\nabla J(s_0)=G(s_0)$,
  \item[iv)] $\nabla G(0)=P$.
\end{itemize}
From the last two conditions, note that
\begin{equation}
\nabla J(0)=G(0)=0,~~\nabla^2 J(0)=\nabla G(0)=P>0.
\end{equation}
Therefore, in some neighborhood of $s_0=0$, the cost functional $J(s_0)$ can be bounded by a convex function of $s_0$, that is,
\begin{equation}
\exists~ k>0,~~\mbox{such that}~~J(s_0)\leq k|s_0|^2. \label{ineq:bound_J}
\end{equation}
This result will be employed in the following subsection.

\subsection{Optimal control for the nonlinear system}

Recalling that $u=B_2^*p$, the cost functional $J(\cdot)$ is
\begin{equation}
J(s)=\int\limits_t^{+\infty}\left( |Cs|^2+|B_2^*p|^2-\gamma^2|w|^2 \right)dt,
\end{equation}
where $s(t)$ is taken as the initial value. Taking the time derivative gives
\begin{eqnarray}
\frac{d}{dt}J(s)&=&\left\langle \nabla J(s), \dot{s} \right\rangle=\left\langle G(s), \dot{s} \right\rangle \nonumber \\
&=&\langle G(s),B_1w\rangle-\left\langle G(s), D_sAs+F(s)+B_2B_2^*G(s) \right\rangle \nonumber \\
&=&\langle B_1^*G(s),w\rangle-\frac{1}{2\gamma^2}|B_1*G(s)|^2-\frac{1}{2}|B_2^*G(s)|^2-\frac{1}{2}|Cs|^2 \nonumber \\
&\leq& \frac{\gamma^2}{2}|w|^2+\frac{1}{2\gamma^2}|B_1*G(s)|^2-\frac{1}{2\gamma^2}|B_1*G(s)|^2-\frac{1}{2}|B_2^*G(s)|^2-\frac{1}{2}|Cs|^2 \nonumber \\
&=& \frac{\gamma^2}{2}|w|^2-\frac{1}{2}|B_2^*G(s)|^2-\frac{1}{2}|Cs|^2.
\end{eqnarray}
Integrating the inequality from $0$ to $+\infty$ yields
\begin{equation}
J(+\infty)-J(s_0)< \frac{\gamma^2}{2}\int\limits_0^{+\infty}|w|^2dt-\frac{1}{2}\int\limits_0^{+\infty}\left( |Cs|^2+|B_2^*G|^2 \right)dt.
\end{equation}
Based on the fact that $J(+\infty)=0$ and the relation (\ref{ineq:bound_J}),
\begin{equation}
\frac{1}{2}\int\limits_0^{+\infty}\left( |Cs|^2+|B_2^*G|^2 \right)dt-\frac{\gamma^2}{2}\int\limits_0^{+\infty}|w|^2dt
< k|s_0|^2.
\end{equation}
When calculating the $H^{\infty}$ problem, the initial condition $s_0$ needs to be set to zero; therefore, 
\begin{equation}
\int\limits_0^{+\infty}\left( |Cs|^2+|B_2^*G|^2 \right)dt<\gamma^2\int\limits_0^{+\infty}|w|^2dt.
\end{equation}
Hence, the $H^{\infty}$ norm of the mapping from the disturbance $w(t)$ to the output of the system is less than $\gamma$, just as in the linear case.

\subsection{Stability analysis for the nonlinear system with no disturbances}

Now the stability analysis for the original nonlinear state equations is the only remaining issue. The objective here is to show that the norm of $s(t)$ in the space $L^2(0,\infty; L^2(\Omega))$ is bounded.

Consider the nonlinear equation without disturbances but with feedback control $u = -B_2^* G(s)$
\begin{equation}
\dot{s}+D_sAs+F_0s+F_N(s) +B_2B_2^*G(s)=0. \label{eq:nonlinear_w0}
\end{equation}
Because it is assumed that the pair $(D_sA+F_0, C)$ is detectable, then there exists a bounded operator $D$, such that
$-(D_sA+F_0+DC)$ generates an exponentially stable semigroup. Rewrite equation (\ref{eq:nonlinear_w0}) as
\begin{equation}
\dot{s}+(D_sA+F_0+DC)s=DCs -B_2B_2^*G(s)-F_N(s).
\end{equation}
Then there exists a constant $b>0$ such that
\begin{equation}
|s(t)|\leq e^{-bt}|s_0|+\int\limits_0^t  e^{-b(t-r)}|DCs-B_2B_2^*G(s)-F_N(s)| dr.
\end{equation}
Integrating from $0$ to $+\infty$ and applying Young's inequality for convolution yields
\begin{equation}
\|s(t)\|^2=\int\limits_0^{+\infty}|s(t)|^2dt\leq C|s_0|^2+C\int\limits_0^{+\infty}\left( |Cs|^2+|B_2^*G|^2+|F_N(s)|^2 \right)dt.
\end{equation}
Applying relation (\ref{ineq:bound_J}) and substituting $w(t)=0$ gives
\begin{equation}
\|s(t)\|^2\leq (C+k)|s_0|^2+C\int\limits_0^{+\infty}|F_N(s)|^2dt. \label{ineq:estimate_s_1}
\end{equation}
If $(s,p)\in\Sigma_{\mu}$, that is, $\| s(t) \| \leq \mu$, the following quadratic estimate for $|F_N(s)|^2 = \left| \frac{c_2}{S} s^2 \right|^2$ holds:
\begin{equation}
|F_N(s)|^2\leq C|s(t)|^4\leq C\mu^2|s(t)|^2.
\end{equation}
Substituting this inequality into equation (\ref{ineq:estimate_s_1}) gives
\begin{equation}
(1-C\mu^2)\|s(t)\|^2\leq (C+k)|s_0|^2.
\end{equation}
If it is then required that
\begin{equation}
1-C\mu^2>0,
\end{equation}
in which case
\begin{equation}
\mu<\frac{1}{\sqrt{C}}.
\end{equation}
This provides the following sufficient condition on the solution
\begin{equation}
\|s(t)\|^2\leq \frac{C+k}{1-C\mu^2}|s_0|^2<\infty,
\end{equation}
from which we can conclude that
\begin{equation}
\lim\limits_{t\rightarrow +\infty}|s(t)|=0.
\end{equation}
Therefore, the nonlinear system is locally asymptotically stable.


\section{Conclusions}

A nonlinear partial differential reaction-diffusion equation is used to describe the evolution of financial distress in financial networks.  The equation includes a diffusion term capturing the distress propagation and a reaction term capturing local amplification of stress level.  Because $s(x) = 0$ corresponds to the case with zero stress level, it is sought to stabilize the system about this state.  A stabilizing feedback controller of the corresponding linearized system is constructed based on the solution to the algebraic Riccati equation, and norm estimation of the linearized system with forcing terms is obtained via semi-group theory.  Based on this norm estimation, unique existence of the solution to the controlled nonlinear system equation is proved by showing that the solution is a fixed point of a contraction mapping.  Finally, the nonlinear state equation describing financial distress propagation and amplification is then stabilized about $s(x) = 0$ via a feedback controller given by the solution of the Hamilton-Jacobi equation.

The resulted control $u(x, t)$ is a state feedback that depends on financial distress diffusion coefficient and the nonlinear local amplification term. In this setting, the distress levels are measured spatially (across different geographical regions or across sectors), and the spatial distribution of control input is computed at government or regulator level. Then local value of control input $u(x)$ for a given time $t$ is sent to location $x$ to execute locally. The spatial distribution of the control input $u(x, t)$ also helps to identify the sub-areas like a specific region or sector that needs financial intervention most.
The local control input mapped to real world can take the forms of Bank-specific liquidity assistance or recapitalization programs, targeted purchase programs for specific markets, emergency credit facilities directed at key sectors, etc. With the collaboration of all local decentralized executions of control inputs $u(x, t)$, the overall financial system will achieve stability as a whole. This approach fits well into the new modern financial policy making paradigm where global financial goals are achieved by the synergy of multiple policy-guided decentralized efforts.  

Beyond the applications in financial distress control, because equation (\ref{eq:state_introduction}) is a generic reaction-diffusion equation, with the reaction being modeled as a nonlinear logistic growth term, the presented results may suggest approaches to be applied to similar mathematical models in other physical applications.

\bibliographystyle{unsrt}
\bibliography{finance_ref}

\end{document}